\documentclass[a4paper,11pt]{article}
\usepackage{jinstpub} 

\usepackage{siunitx}
\usepackage[version=4]{mhchem}

\usepackage{xspace}

\proceeding{8$^{\text{th}}$ International Conference on Micro-Pattern Gaseous Detectors\\
14-18 October 2024\\
Hefei, China}

\title{\boldmath Development of a high-rate capable DLC-RPC based on a current evacuation pattern}

\author[a, *]{M.~Takahashi,\note{Corresponding author.}}
\author[b]{S.~Ban,}
\author[c]{W.~Li,}
\author[a, \dag]{A.~Ochi,\note{Deceased.}}
\author[b]{W.~Ootani,}
\author[b]{A.~Oya,}
\author[a]{H.~Suzuki}
\author[c]{and K.~Yamamoto}

\affiliation[a]{Department of Physics, Kobe University,\\
                1-1, Rokkodai-cho, Nada-ku, Kobe, 657-8501, Hyogo, Japan}
\affiliation[b]{International Center for Elementary Particle Physics, The University of Tokyo,\\
                7-3-1, Hongo, Bunkyo-ku, Tokyo, 113-0033, Japan}
\affiliation[c]{Department of Physics, The University of Tokyo,\\
                7-3-1, Hongo, Bunkyo-ku, Tokyo, 113-0033, Japan}

\emailAdd{m.takahashi@stu.kobe-u.ac.jp}

\abstract{A Resistive Plate Chamber using Diamond-Like Carbon electrodes (DLC-RPC) has been developed as a background tagging detector in the MEG~II experiment.
The DLC-RPC is planned to be installed in a high-intensity and low-momentum muon beam.
This detector is required to have a detection efficiency above \qty{90}{\%} with four active gaps in the muon beam due to the limitation of the material budget.
In such an environment, the high current flowing through the resistive electrodes causes a voltage drop, which reduces the performance of the DLC-RPC.
This voltage drop can be suppressed by implementing a current evacuation pattern, though discharges are more likely to occur near the pattern.
Therefore the pattern must be covered by a protection cover made of an insulator.
In this study, electrode samples with a current evacuation pattern and different widths of protection cover (\qtylist{0.2; 0.8}{\mm}) have been produced, and their performance and stability were measured.
The detection efficiency of a single-gap chamber for $\beta$-rays from a \ce{^{90}Sr} source was measured to be up to approximately \qty{60}{\%} in both electrode samples.
The target efficiency can be achieved even with a drop of \qtyrange[range-phrase=~--~, range-units=single]{100}{150}{\V}.
On the other hand, after more than a dozen hours of operation, discharges suddenly occurred and the detector was prevented from further operation.
These discharges created current paths on the spacing pillars.
This serious problem must be investigated and solved in the future.}


\arxivnumber{2501.05128} 

\begin{document}
\maketitle
\flushbottom

\section{Introduction}
\label{sec:intro}

A novel type of resistive plate chamber, based on diamond-like carbon (DLC) electrodes~\cite{IEKI2024169375}, is under development as a background tagging detector in the MEG~II experiment~\cite{megii-design}.
This detector should have an ultra-low mass (less than \qty{0.1}{\%~X_0}) and a high-rate capability (up to \qty{3}{\MHz/\cm\squared}), which needs to be operated in a high-intensity (\num{7e7}~particles/s) and low-momentum (\qty{28}{\MeV}/$c$) muon beam.
In addition, a detection efficiency above \qty{90}{\%} is required for minimum ionization particles in an active region of \qty{16}{\cm} in diameter.
The number of active gaps is limited to up to four in order to fulfill the requirement of the material budget.

Reference~\cite{IEKI2024169375} proposed strip-shaped current evacuation patterns to achieve a high rate capability with a scalable detector design.
A small-size prototype chamber of the DLC-RPC was constructed to study this design~\cite{YAMAMOTO2023168450}.
Gas gaps of the DLC-RPC were supported by spacing pillars formed by photolithographic technology, and solder resist was used as a material for the spacing pillars.
In addition, the current evacuation pattern was implemented in this prototype detector.
However, we could not perform any studies with this prototype because the gap thickness was non-uniform due to technical problems of spacing pillar formation using this solder resist.
Recently, we succeeded in improving the gap thickness uniformity by using another solder resist material, which is Dynamask (Ethernal~Tech.~Corp.).
Therefore, a prototype detector that has uniform spacing pillars and a current evacuation pattern was newly fabricated to study this strip structure.
This paper describes the performance and operation stability of the prototype.

\section{DLC-RPC with current evacuation pattern}
\label{sec:strip}

\begin{figure}[htbp]
    \centering
    \includegraphics[width=.28\textwidth]{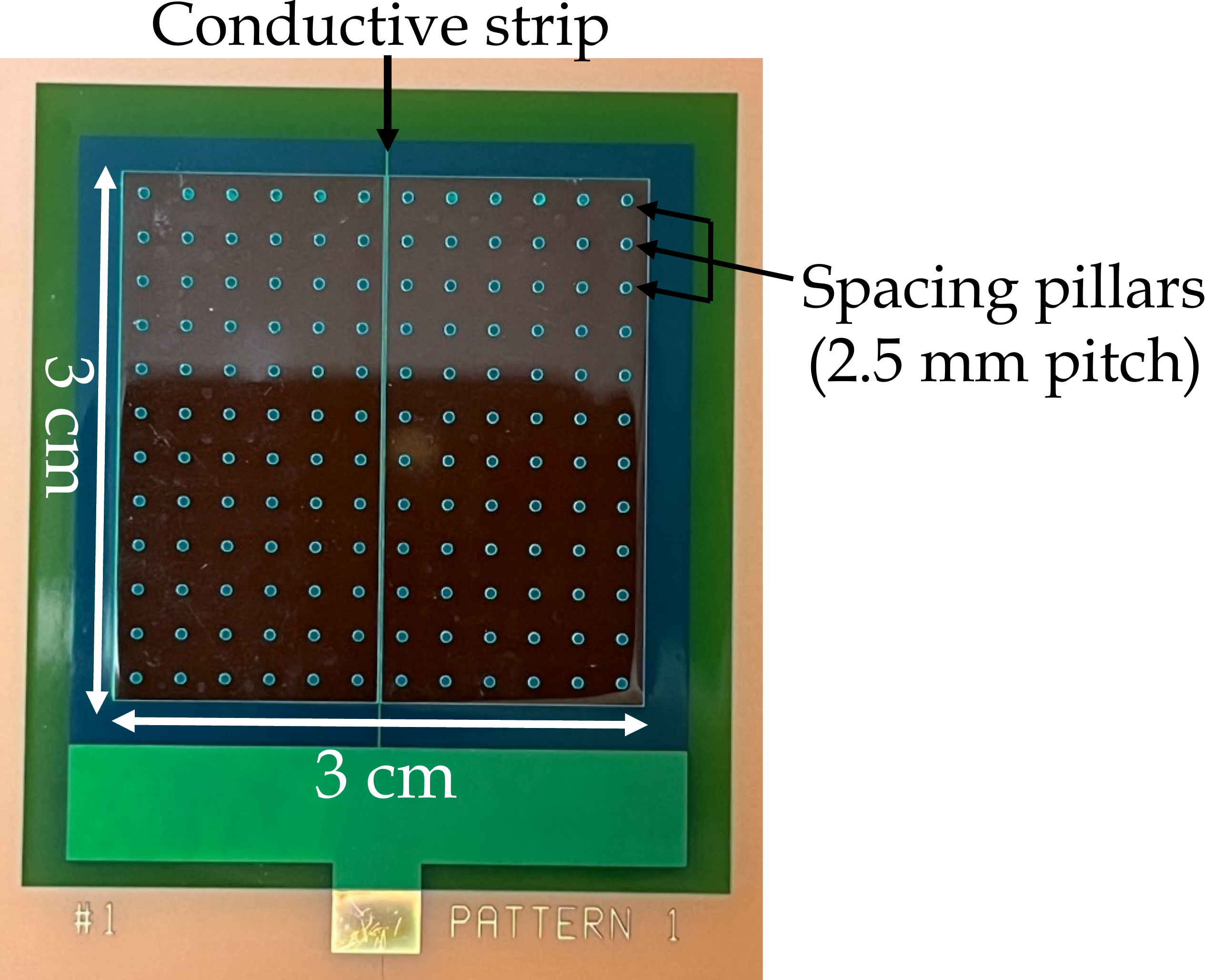}
    \quad
    \includegraphics[width=.26\textwidth]{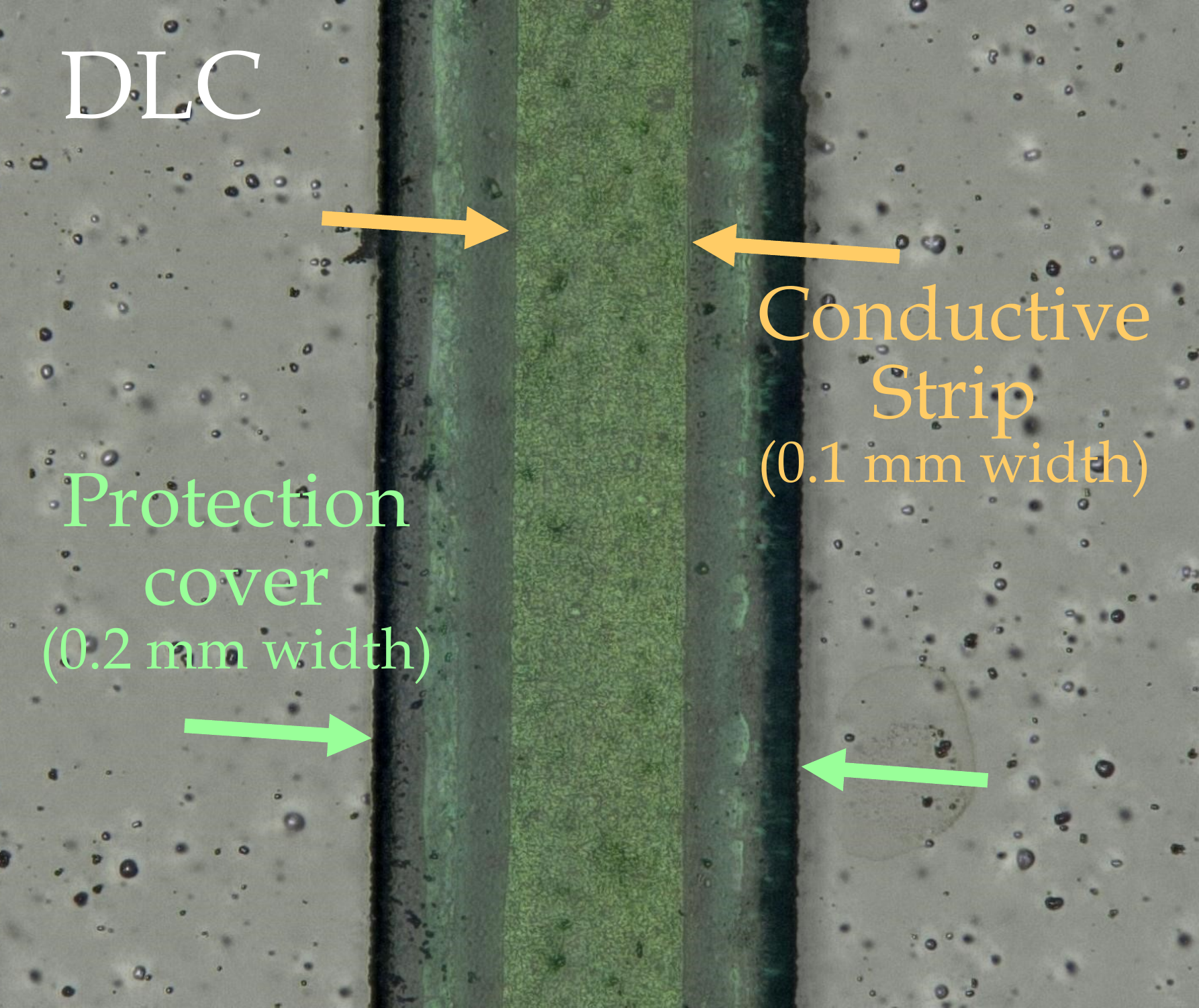}
    \qquad
    \includegraphics[width=.3\textwidth]{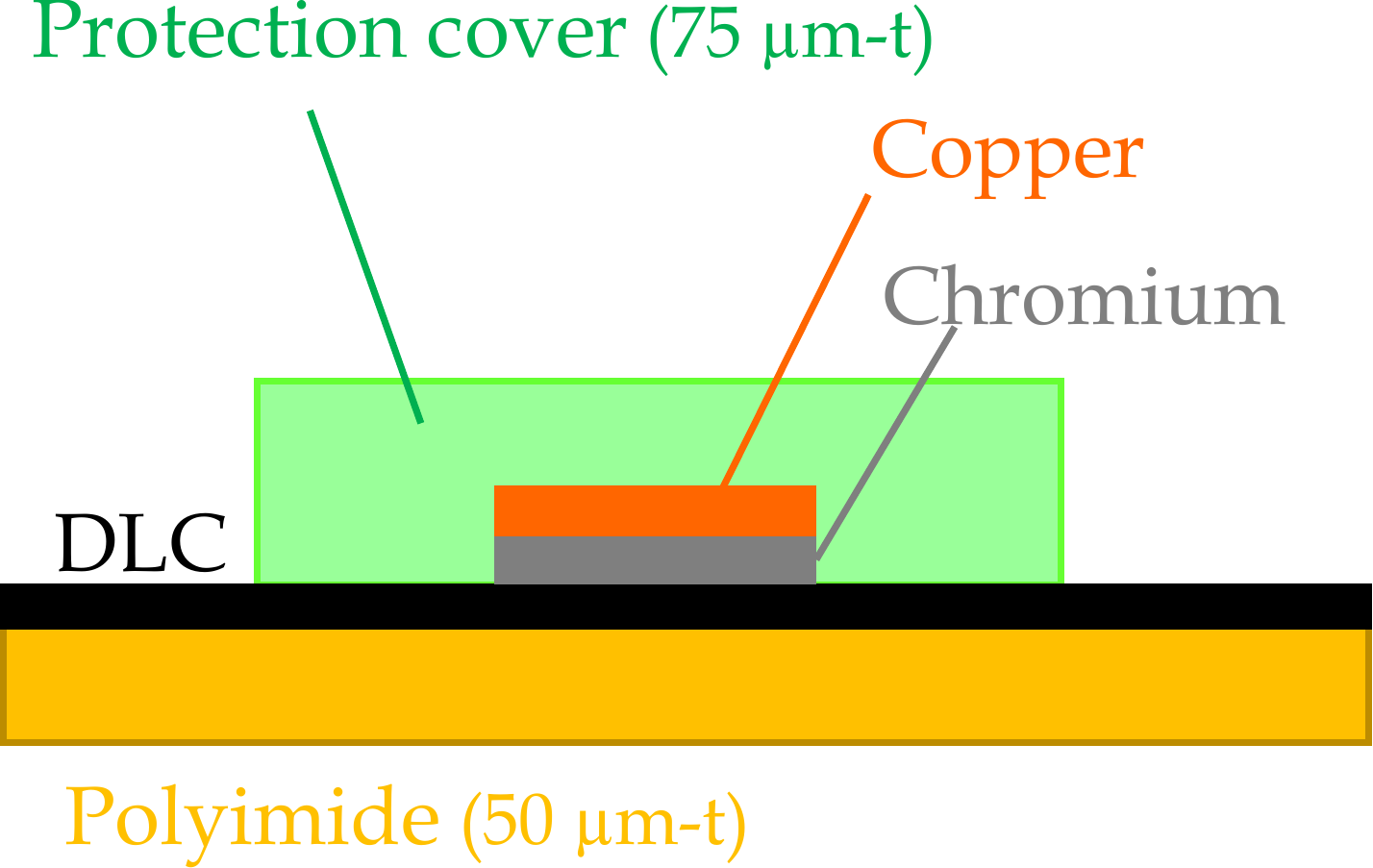}
    \caption{Pictures and a schematic of the DLC-RPC electrode with the current evacuation pattern. (Left) Overview. (Center) Enlarged view of the strip structure. (Right) Cross-section of the strip structure.\label{fig:electrode}}
\end{figure}

Figure~\ref{fig:electrode} shows the DLC-RPC electrode with the current evacuation pattern.
DLC was directly sputtered onto the \qty{50}{\um} thick polyimide film.
The conductive current evacuation pattern was implemented on the DLC layer and covered with a protection layer whose material is the same as the pillars.
This pattern was formed by the lift-off process.
Chromium was sputtered onto the DLC layer, followed by copper sputtering onto the chromium layer.
This sandwich structure is motivated by the good adherence of chromium with DLC.
This electrode is made with pillars of $\sim$\qty{365}{\um}~thickness, \qty{0.6}{\mm} diameter and \qty{2.5}{\mm}~pitch, with a current evacuation pattern of \qty{0.1}{\mm}~width, and a \qtyproduct[product-units=single]{3x3}{\cm\squared} of active region.
The variation of the pillar thickness was measured to be below $10~\upmu\mathrm{m}$.
The surface resistivity of DLC was measured to be \qtyrange[range-phrase=~--~, range-units=single]{6}{15}{\Mohm/sq}.
Different electrode samples with \qtylist{0.2; 0.8}{\mm} widths of the protection cover were tested to investigate the width needed to avoid discharge.
According to the results and discussions in Reference~\cite{IEKI2024169375}, the voltage drop will be down to \qty{100}{\V} at the expected operation voltage in the muon beam for this configuration.

\section{Experimental setup}

\begin{figure}[htbp]
    \centering
    \includegraphics[width=.9\textwidth]{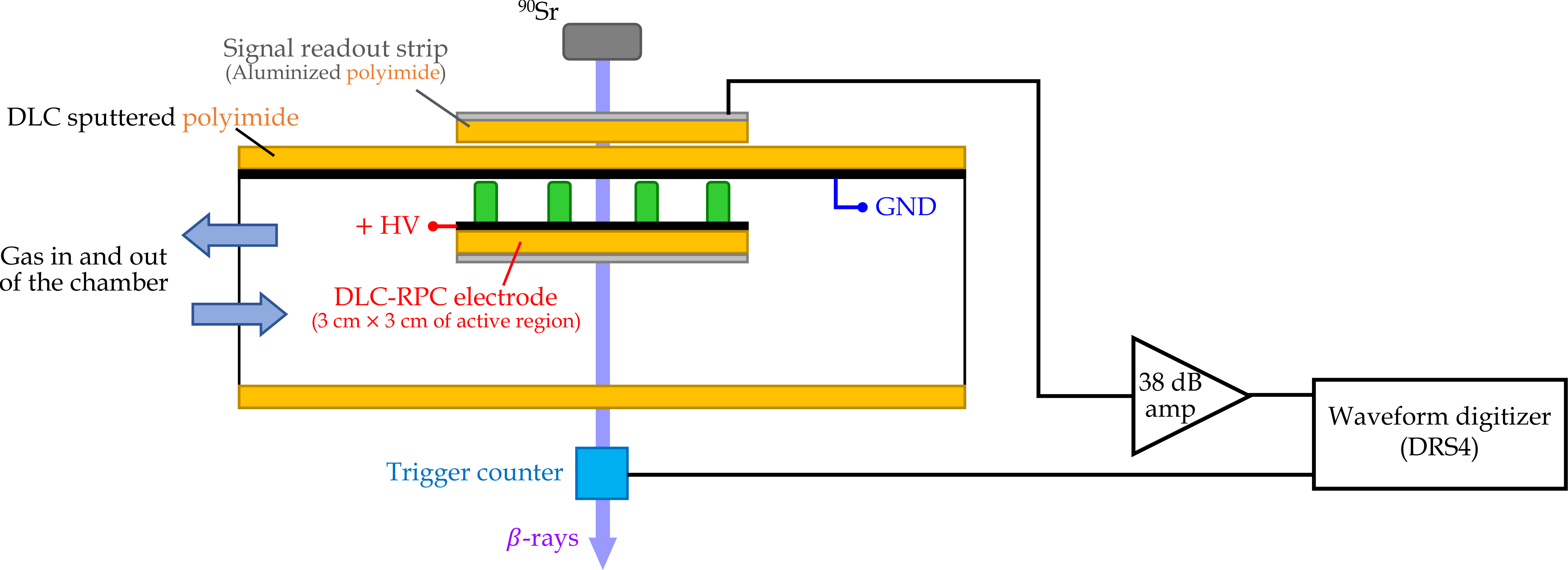}
    \caption{Experimental setup. The DLC-RPC electrode is placed in a chamber filled with working gas.\label{fig:setup}}
\end{figure}

In this study, we measured the detection efficiency for $\beta$-rays and assessed the stability of the DLC-RPC.
An experimental setup is shown in Figure~\ref{fig:setup}.
The electrode shown in Figure~\ref{fig:electrode} was used as an anode electrode.
A bare DLC sputtered polyimide was used as a cathode electrode, in which the surface resistivity of DLC was measured to be \qtyrange[range-phrase=~--~, range-units=single]{10}{15}{\Mohm/sq}.
These electrodes were placed facing each other to form the DLC-RPC gas gap.
This gap was filled with \ce{C2H2F4 / SF6 / iC4H10} ($94 / 1 / 5$).
Trigger counter signals were fed directly into a DRS4 waveform digitizer~\cite{RITT2004470}, and detection signals of the DLC-RPC were induced in an aluminum strip outside the gas gap, amplified by \qty{38}{\dB}, and then fed into the DRS4.
In this measurement, a detection efficiency of the DLC-RPC was calculated as the percentage of events in which the DLC-RPC signal exceeded a \qty{20}{\mV} signal threshold, which corresponds to a \qty{25}{\femto\coulomb} charge threshold in the readout strip among events triggered by a trigger counter.

\section{Results and Conclusion}

\begin{figure}[htbp]
    \centering
    \includegraphics[width=.44\textwidth]{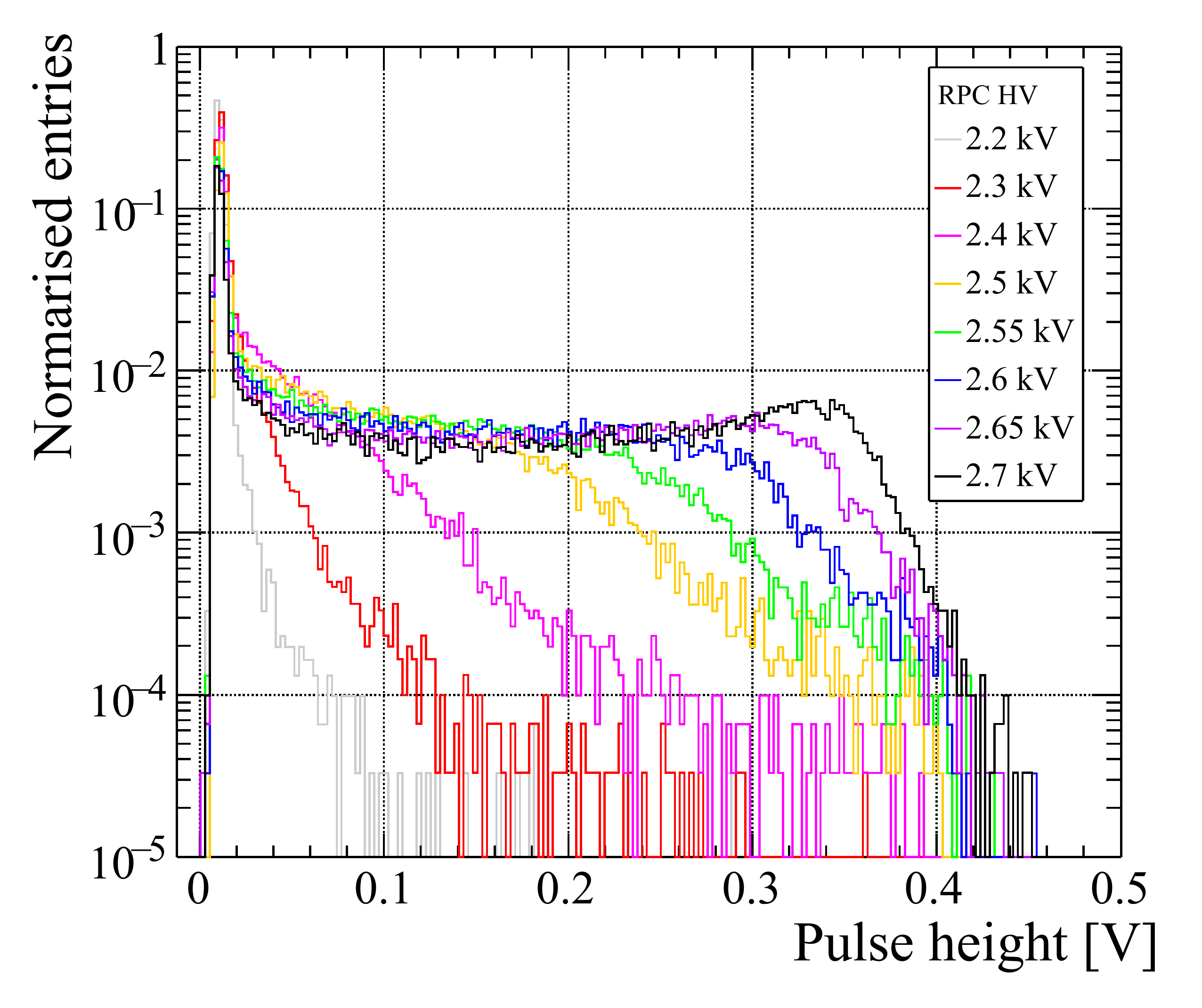}
    \qquad
    \includegraphics[width=.41\textwidth]{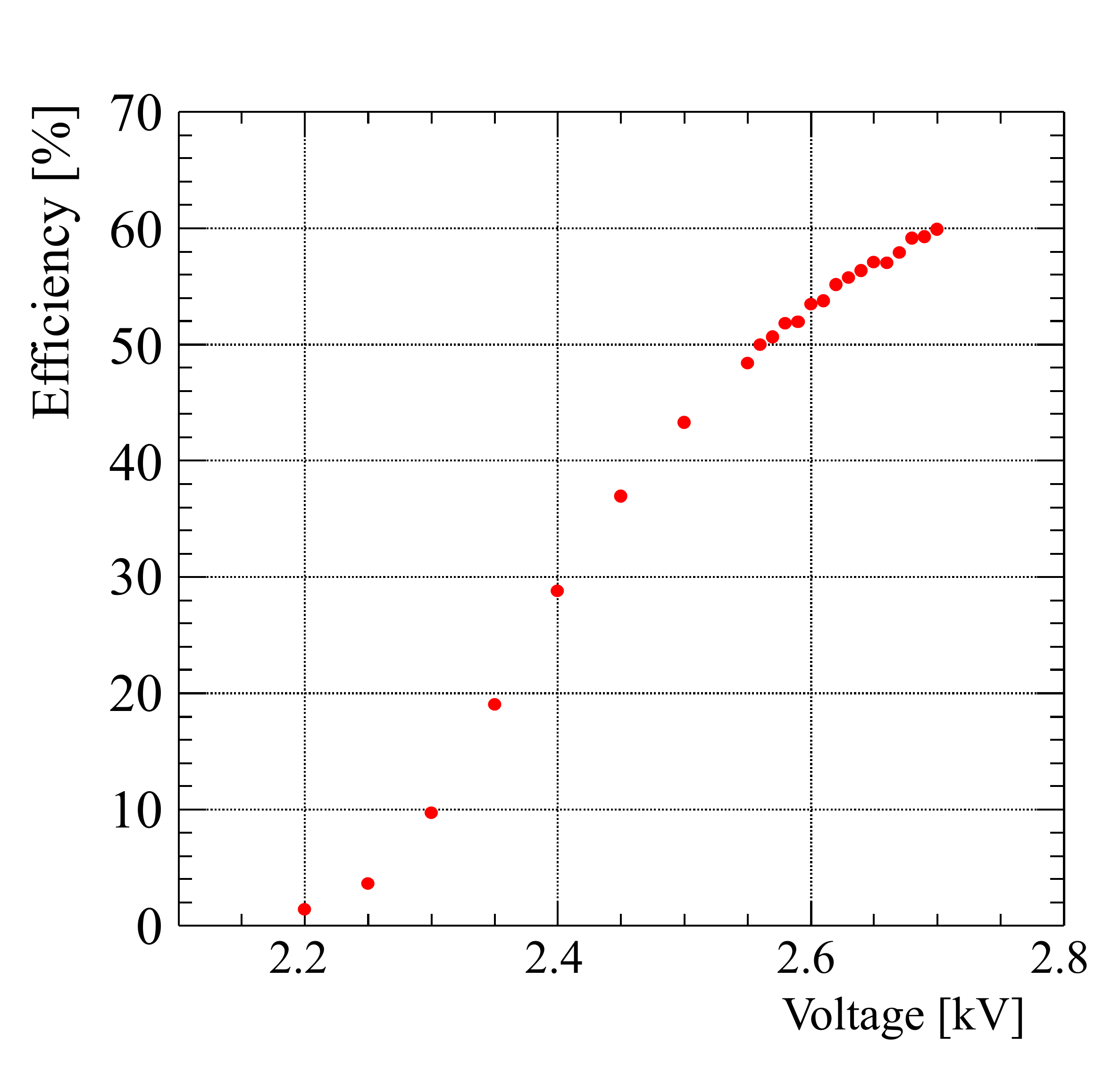}
    \caption{Performance of the DLC-RPC for the $\beta$-rays. (Left) Pulse height spectra of the DLC-RPC triggered by a trigger counter. (Right) Detection efficiency. 
    \label{fig:performance}}
\end{figure}

Figure~\ref{fig:performance} shows the performance of the DLC-RPC for $\beta$-rays from a \ce{^{90}Sr} source.
In this measurement, the $\beta$-rays were collimated to a beam of \qty{2}{\mm} diameter, which was irradiated far from the conductive pattern to avoid instability due to the strip structure.
This instability is attributed to an excessive gas avalanche process caused by the distortion of an electric field caused by the protection cover and a charging up of it, as well as a quenching capability against discharges is low due to the short distance over which the charge flows through the resistive layer.
The measured hit rate was $\mathcal{O}\left(\qty{1}{kHz}\right)$.
The $n$-gaps efficiency can be estimated with the single-gap efficiency $\epsilon_{1}$ by $\epsilon_{n}=1-(1-\epsilon_{1})^{n}$.
The measured efficiency of \qty{60}{\%} at \qty{2.75}{\kV} is enough to reach a target efficiency of \qty{90}{\%} with four gaps.
A minimum single-layer efficiency to obtain this target efficiency is about \qty{45}{\%}, which requires an operating voltage of around \qty{2.5}{\kV} from the right side of Figure~\ref{fig:performance}.
Since the operating voltage reached \qty{150}{\V} higher than this minimum required voltage, we can expect that the detection efficiency will not fall below \qty{45}{\%} (corresponds to \qty{90}{\%} efficiency with four layers) even with a \qty{100}{\V} voltage drop in the muon beam as described in Section~\ref{sec:strip}.

Next, we investigated the maximum voltage that can be applied to the gap for different widths of the protection cover (\qtylist{0.2; 0.8}{\mm}).
In this measurement, the whole active region was irradiated with non-collimated $\beta$-rays with a hit rate of $\mathcal{O}\left(\qty{100}{\kHz}\right)$.
For both widths of the protection cover, the efficiency requirement was fulfilled although the maximum operation voltages were reduced by \qtyrange[range-phrase=~--~, range-units=single]{50}{60}{\V} due to the high current flowing irregularly around the strip structure.
Increasing the cover width can slightly suppress a frequency of such irregular currents, though it also increases the dead area.
In practice, since the chamber needs to be operated in an even more intense muon beam, the width of the protection cover will be optimized to account for safety factors and its occupancy.

\begin{figure}[htbp]
    \centering
    \includegraphics[width=.45\textwidth]{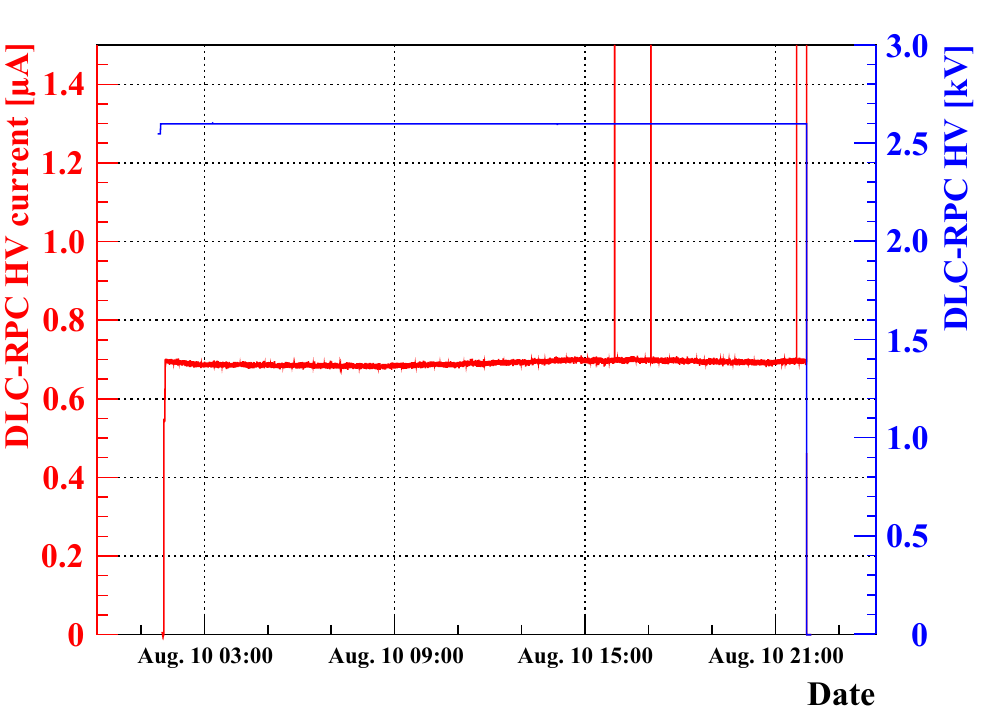}
    \qquad
    \mbox{\raisebox{5mm}{\includegraphics[width=.3\textwidth]{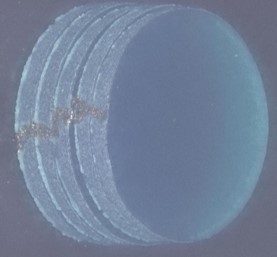}}}
    \caption{Status of the long-term operation. (Left) Operation voltage and measured current versus time. (Right) Picture of the damaged spacing pillar.\label{fig:status}}
\end{figure}

Finally, the long-term stability was assessed.
The whole region was continually irradiated with non-collimated $\beta$-rays with a hit rate of $\mathcal{O}\left(\qty{100}{\kHz}\right)$.
The left plot in Figure~\ref{fig:status} shows the operation history.
The operating voltage was set to \qty{2.6}{\kV}, which is \qtyrange[range-phrase=~--~, range-units=single]{30}{50}{\V} lower than the target voltage.
Discharges suddenly occurred after about \qty{14}{hours}, and the DLC-RPC eventually tripped after \qty{20}{hours}.
The discharge created a current path on one of the pillars (the right in Figure~\ref{fig:status}) and prevented the operation of the chamber.
Furthermore, higher hit rates tended to make the chamber more unstable, but we need further investigation in this dependence.

Summarizing, the measured performance looks promising for the prototype in terms of the requirement for detection efficiency.
On the other hand, the long-term instability, caused by materials inside gaps, is problematic in the MEG~II experiment.
Since low surface resistivity can cause such instability, it should be adjusted to safer ranges (above \qty{20}{\Mohm/sq}).
Besides, the avalanche charge could gradually burn the polymer in the pillars, eventually leading to the discharges as shown in Figure~\ref{fig:status}.
We need further investigation into causes and solutions to suppress damage to the material.
Another scenario that caused instability is an ionic contamination of residuals from a resist-dissolving process.
As mentioned in Section~\ref{sec:intro}, spacing pillars were formed by photolithographic technology.
In this process, the resist is dissolved with a chemical solution after curing by exposure, and a residue from the dissolving process could become a current pathway.
We will plan to add an electrode cleaning process to address this contamination.

\acknowledgments

We are grateful for the technical support provided by the Micro-Pattern Technology workshop at CERN and for the technical advice of Rui De Oliveira.
This work is supported by JSPS, Japan KAKENHI Grant Number JP21H04991, JST SPRING, Japan Grant Number JPMJSP2148.


\bibliographystyle{JHEP}
\bibliography{biblio}

\providecommand{\href}[2]{#2}\begingroup\raggedright\begin{thebibliography}{1}

\bibitem{IEKI2024169375}
K.~Ieki et~al., \emph{{Prototype study of $0.1\%\,X_0$ and $\mathrm{MHz/cm^2}$ tolerant Resistive Plate Chamber with Diamond-Like Carbon electrodes}}, \href{https://doi.org/https://doi.org/10.1016/j.nima.2024.169375}{\emph{NIM~A} {\bfseries 1064} (2024) 169375}.

\bibitem{megii-design}
A.M.~Baldini et~al., \emph{{The design of the MEG II experiment}}, \href{https://doi.org/10.1140/epjc/s10052-018-5845-6}{\emph{EPJ~C} {\bfseries 78} (2018) }.

\bibitem{YAMAMOTO2023168450}
K.~Yamamoto et~al., \emph{{Development of ultra-low mass and high-rate capable RPC based on Diamond-Like Carbon electrodes for MEG~II experiment}}, \href{https://doi.org/https://doi.org/10.1016/j.nima.2023.168450}{\emph{NIM~A} {\bfseries 1054} (2023) 168450}.

\bibitem{RITT2004470}
S.~Ritt, \emph{{The DRS chip: cheap waveform digitizing in the GHz range}}, \href{https://doi.org/https://doi.org/10.1016/j.nima.2003.11.059}{\emph{NIM~A} {\bfseries 518} (2004) 470}.

\end{thebibliography}\endgroup






\end{document}